\documentstyle[fleqn,espcrc2,epsfig]{article}
\title{$\Omega^-$ Hyperon Decay Modes with the
HyperCP Experiment at Fermilab.}
\author{Prof. Nickolas Solomey
\address{Illinois Institute of Technology, Chicago, USA}}

\begin{document}

\begin{abstract}
The HyperCP experiment (Fermilab E871) has the largest recorded sample of
$\Omega^-$ hyperon decays, allowing new searches for three rare decay
modes, each with five charged tracks in the final state. Results are
presented for $\Omega^-\to\Xi^-\pi^+\pi^-$; $\Omega^- \to
\Xi^0(1530)\pi^-$; and the flavor-changing neutral-current decay $\Omega^-
\to \Xi^-\mu^+\mu^-$. Normalization of these measurements were to the
decay $\Omega^- \to \Lambda K^-$ with a five-track topology similar to the
signal mode with the subsequent decay of $\Lambda \to p \pi^-$ and $K^-
\to \pi^- \pi^+ \pi^-$ in the decay volume.
\end{abstract}

\maketitle

\section{Theoretical Background}
The $\Omega^-$ hyperon with three strange quarks is of great theoretical
interest because of its unique ability to tell us about the processes
involved in the decay \cite{eilam}. However to be able to actually use
this wonderful tool provided by nature it is first necessary to obtain a
large sample of interesting decays. The HyperCP experiment has made great
progress in this pursuit with a recent observation of a rare decay mode.

The decays of interest and their theoretical expected \cite{singer} or
previously reported Branching Ratios (BR) are:
\begin{eqnarray}
\Omega^- & \rightarrow & \Xi^- \gamma \hspace{1.2cm} 1.6\times10^{-4} \\
    & \rightarrow & \Xi^- \pi^+ \pi^- \hspace{.5cm} 4.3\times10^{-4}
\hspace{0.3cm}
{\rm \cite{wa2}} \\
    & \rightarrow & \Xi^- e^+ e^- \hspace{.6cm} 1.4\times10^{-6}  \\
    & \rightarrow & \Xi^- \mu^+ \mu^- \hspace{.55cm} 6.6\times10^{-8}
\end{eqnarray}
(The equivalent anti-particle processes are also of interest.) 
All of these
decay modes have a high-$Q$ released energy and should not be limited by
phase space. The radiative decay mode listed first has yet to be seen, but
it has a high theoretical BR. The second decay mode has many
possible suppressed feynman diagrams that can contribute (such as tree,
penguin, and double-Zweig) and perhaps others due to new particles
including higgs or supersymmetry, but experimentally it has a substantial
BR. The last two decays are flavor-changing neutral currents,
which have never been seen with hyperon decays.

Observation of these decays, especially with high statistics, can be a
useful tool for both probing the accuracy of the standard model, and as a
means to look for new physics. A large decay sample can be useful for more
than just the BR measurement. The subsequent decay of the
$\Xi^-$ can tell us about the decay itself by using its analyzing power to
look for an asymmetry in the decay \cite{hara}, similar to that done with
other hyperon radiative decays \cite{ktevrad}.

There are also two other processes that can contribute to the decay in
equation (2). These are when the final observed five tracks have gone
through a short lived resonance:
\begin{eqnarray*}
\Omega^- & \rightarrow & \Xi^0(1530) \hspace{0.15cm}   \pi^-
\hspace{3.5cm} (5)\\
& &   \hspace{0.24cm}  \rightarrow  \Xi^-  \pi^+  \\
    &  \rightarrow & \Xi^- \sigma^0(400)  \hspace{3.85cm}  (6) \\
& &  \hspace{0.65cm} \rightarrow \pi^+ \pi^-
\end{eqnarray*}
Observation of the $\sigma^0(400)$ would be important since the existence
of this state is highly debated and has never been observed in any decay
chain. Its mass value is stated in the PDG as being from 400 to 1200
Mev/c$^2$ and its Breit-Wigner width is anywhere from 600 to 1000
MeV/c$^2$! All of these parameters could be resolved if this particle is
studied in a decay.

\section{New Results in $\Omega^-$ Decays}
\begin{figure}
\begin{center}
\mbox{
\epsfig{file=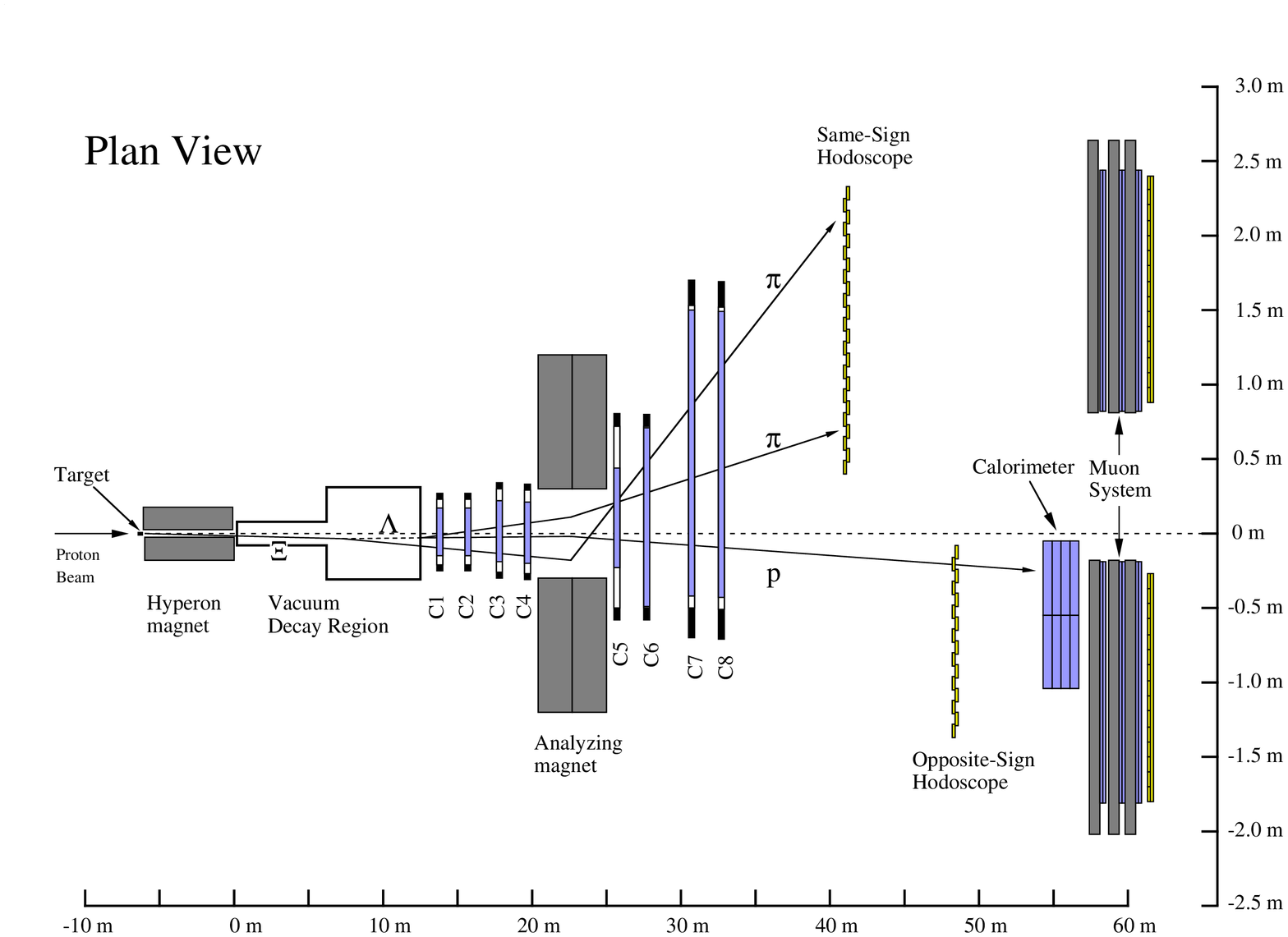,%
height=0.725\linewidth}
}
\end{center}
\caption{A schematic depiction of the E871 HyperCP experiment at
Fermilab.} \label{e871}
\end{figure}

The HyperCP experiment E871 at Fermilab took data in 1997 and 1999
\cite{e871}. The experiment is depicted in figure \ref{e871}. It was
designed to search for CP violation in Hyperons \cite{pi}. The 800
GeV/c proton beam hit a target at the entrance of a hyperon channel with a
nominal momentum of 170 GeV/$c$ and a bite of $\pm 35$ GeV/$c$ FWHM. The
experiment's decay volume started at the exit of the hyperon channel and
was 13 meters long. The experiment itself is a charged particle tracking
spectrometer, with 4 MWPC before a high-field analysis magnet and 4 MWPC
after the analysis magnet. These were followed by a hodoscope of 10-cm
wide vertical paddles in the positive and negative arms of the
spectrometer. There was a muon identification system further downstream
that consisted of a muon hodoscope and three layers of steel and
proportional tubes.

The trigger of the experiment consisted of at least two charged tracks of
opposite polarity and energy in the calorimeter, which was a combination
electromagnetic and hadronic calorimeter. The experiment wrote more than
40,000 5Gb Exabyte data tapes. These raw data tapes were processed to
produce different output data streams which had further requirements such
as three charged tracks, good masses, and identified muons depending upon
the intended physics use of each stream. Anti-particle processes (referred
to as ``positive'' data) were studied in the same spectrometer under
identical conditions (except for their production momentum spectra and
daughter-particle interaction cross sections) by reversing the magnetic
field in the hyperon channel and analysis magnets.

The results presented here come from an analysis of the stream with at
least three charged tracks for the decay mode $\Omega^- \rightarrow \Xi^-
\pi^+ \pi^-$ and at least three charged tracks plus two opposite polarity
muon tracks for the decay mode $\Omega^- \rightarrow \Xi^- \mu^+ \mu^-$.
The selection cuts were devised with a blind analysis on signal,
normalizing and background modes generated with a dedicated Monte-Carlo
simulation program. The cuts used in this analysis were: (a) three
negative and two positive tracks, (b) all decay vertices inside the decay
volume, (c) vertex topology consistent with the intended decay, (d) total
$\Omega^-$ momentum between 120 and 220 GeV/$c$, (e) all reconstructed
masses within $\pm 3 $ sigma of their nominal mass, and (f) the
reconstructed $\Omega^-$ track appeared to come from the aperture of the
hyperon channel exit. For the muon decay mode of equation (4) an
additional requirement from the muon identification system was imposed.
For the $\overline{\Omega}$, the number of different charged particles
required was inverted.

Normalization of the candidate events could be done with both the five-
and three-track decays $\Omega^- \rightarrow \Lambda K^-$ where the $K^-$
either does or does not decay into three charged pions. In either case the
$\Lambda$ decays into proton and $\pi^-$. The five-track decay mode is the
most valuable decay for normalization since the topology is very similar
to the signal modes and is the one used in all BR calculations presented
here. The Monte-Carlo was used for acceptance corrections among the
various modes. For the signal mode of equation (2) a uniformly populated
phase
space generator was used. The Monte-Carlo was also used to simulate
possible
backgrounds before opening the signal box. At that stage of the analysis
there was confidence that no significant backgrounds 
to these decay modes existed.

\section{Observations and Measurements}
\begin{figure}[t]
\begin{center}
\mbox{
\epsfig{file=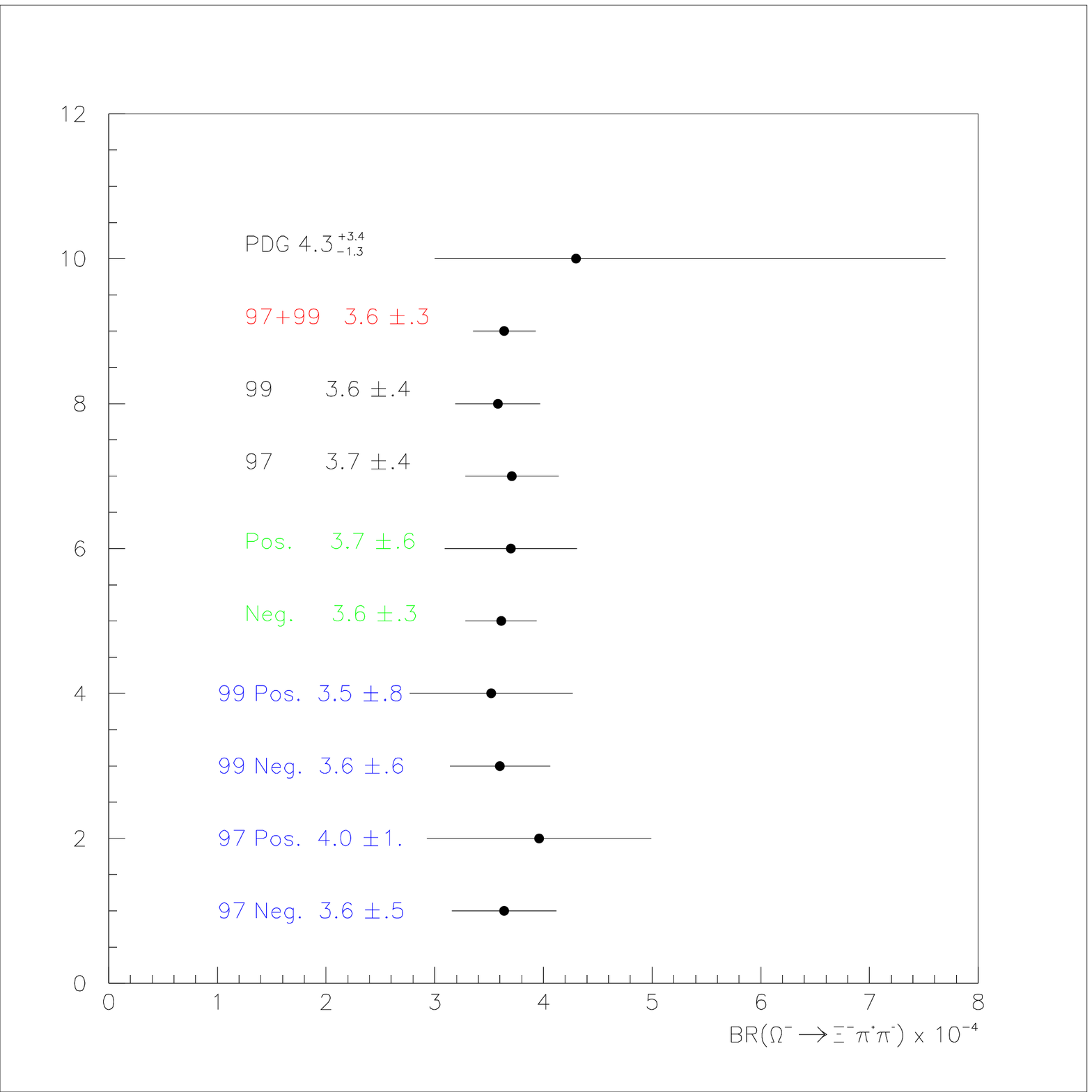,%
height=0.98\linewidth}
}
\end{center}
\caption{Preliminary BR for the $\Omega^- \rightarrow \Xi^-
\pi^+ \pi^-$ for all four data samples and various combinations. Errors
are statistical only.}
\label{br}
\end{figure}

The full HyperCP data sample was analyzed \cite{solomey}. The observed
``negative'' five-track signal of the decay $\Omega^- \rightarrow \Xi^-
\pi^+ \pi^-$ is shown in figure \ref{data}, and the normalization was done
with the five-track decay mode $\Omega^- \rightarrow \Lambda K^-$ where
$K^- \rightarrow \pi^- \pi^+ \pi^-$ and $\Lambda \rightarrow p \pi^-$
shown in figure \ref{data2}.
{\begin{figure}[t]
\begin{center}
\mbox{
\epsfig{file=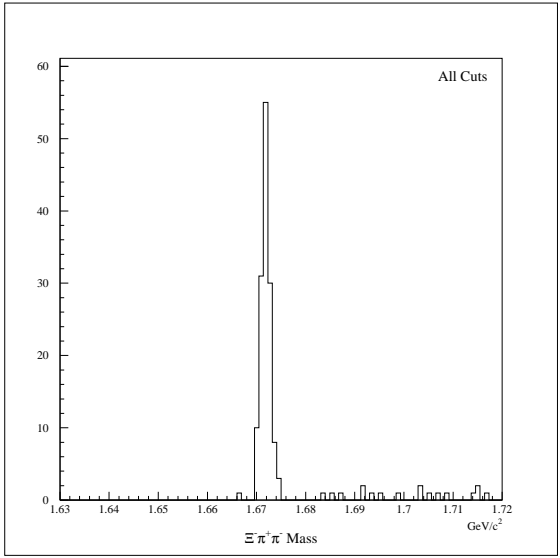,%
height=0.98\linewidth}
}
\end{center}
\caption{The observed events for the five-track signal mode of equation
(2) is shown with all of the negative hyperon beam data.}
\label{data}
\end{figure}
\begin{figure}[t]
\begin{center}
\mbox{
\epsfig{file=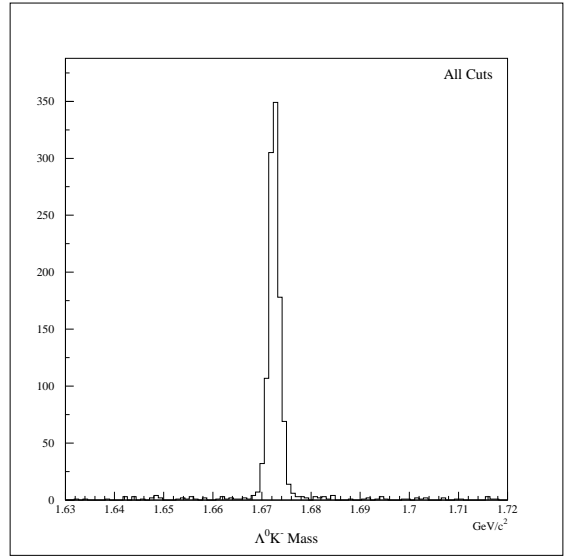,%
height=0.98\linewidth}
}
\end{center}
\caption{The observed events for the five-track normalizing mode $\Omega^-
\rightarrow \Lambda K^-$ where the $K^-$ decays into three charged pions
and the $\Lambda$ into $p \pi^-$ in the experiment's decay volume before
the first MWPC, for all of the negative hyperon beam data.}
\label{data2}
\end{figure}
} When combined with the acceptance correction from Monte-Carle simulation
and the appropriate known BR from the PDG \cite{pdg} it permitted
measuring
\begin{eqnarray*}
BR(\Omega^{-} \to \Xi^{-} \pi^{+} \pi^{-}) & = & \hspace{2.85cm} (7)\\
& & \hspace{-3.2cm} \frac{N_{\Omega^{-} \to \Xi^{-} \pi^{+} \pi^{-}}}
{N_{\Omega^- \to \Lambda K3pi^-}}
\frac{A_{\Omega^- \to \Lambda K3pi^-}}
{A_{\Omega^{-} \to \Xi^{-} \pi^{+} \pi^{-}}} \times \\
& & \hspace{-3.2cm} \frac{BR(\Omega^- \to \Lambda K^-)\times BR(K^- \to
\pi^- \pi^+ \pi^-)} {BR(\Xi^- \to \Lambda \pi^-)}.
\end{eqnarray*}
All data sets and their combinations yields statistically consistent
results as shown in figure \ref{br}. The signal events within $\pm3
$ sigma were used, and the background under the signal peak was estimated
using the side bands at $\pm 7{\rm -}11 $ sigma.

\section{Conclusion}
The final combined preliminary BR for this decay mode is $3.6
\times 10^{-4}$ with a statistical error of $\pm 0.3 \times 10^{-4}$,
which is slightly
better than a 10\% measurement. The HyperCP data sample was also used to
measure the $\overline{\Omega}$ hyperon (referred to in the graph as
``Pos\@.'' data), and this yields a very similar result within errors. The
full data sample was also used to look for the flavor-changing
neutral-current decay modes with muon pairs. No candidate events were
observed yielding a preliminary BR limit with only the
negative data as $< 1 \times 10^{-4}$ at the 90\% confidence level.

\begin{figure}[t]
\begin{center}
\mbox{
\epsfig{file=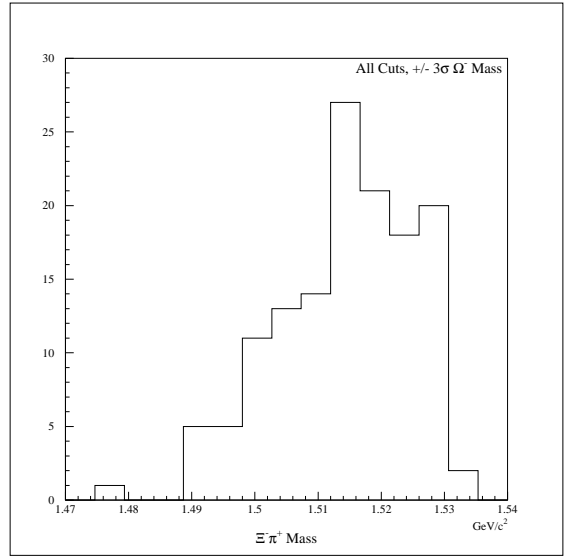,%
height=0.98\linewidth}
}
\end{center}
\caption{Using only those $\Omega^-$ within
$\pm 3 $ sigma of its known mass, the $\Xi^0$(1530) resonance
is not immediately evident.}
\label{data3}
\end{figure}

A first look at the resonance mode of equation (5) shows no clear sign of
its existence (see figure \ref{data3}), but the $\Xi^0(1530)$ is a wide
resonance and spin has not yet been taken into account in the Monte-Carlo
simulation of this decay so the exact expectations of its signature may be
not be an obvious peak.

Further work is needed to refine these results and their associated
systematic errors which are not worse than these statistical
errors. There are also plans to use this data to understand what, if
anything, resonance decays could be contributing. Furthermore, by using
the self-analyzing power of the $\Xi^-$ decays it should be possible to
look for any asymmetries of the $\pi^+ \pi^-$ pair, similar to the
asymmetry seen in hyperon radiative decays, to yield clues to the decay
mechanism.

\section*{Acknowledgments}
The data used in this analysis is the direct result of a lot
of hard work by members of the HyperCP experiment, and by the Fermilab and
University technical staffs. This work was partly done in conjunction with
Nathan Stobie, a senior thesis student at the Illinois Institute of
Technology. I am also greatly indebted to Jon Rosner and Paul Singer for
many useful discussions.

\end{document}